\documentclass[11pt,twoside]{article}

\usepackage{asp2006}
\usepackage{epsf}
\usepackage{lscape}
\usepackage{graphicx}

\newcommand{\Hi}{\textup{H\,{\mdseries\textsc{i}}}}
\newcommand{\HI}{\textup{H\,{\mdseries\textsc{i}}}~}

\def\kms{{km~s$^{-1}$}}

\def\arcsec{\hbox{$^{\prime\prime}$}}

\begin{document}
\title{Centaurus A: Morphology and kinematics of the HI disk}   
\author{C. Struve$^{1,2}$, R. Morganti$^{1,2}$, T.A. Oosterloo$^{1,2}$}   
\affil{$^1$Netherlands Foundation for Research in Astronomy, P.O. Box 2, 7990 AA Dwingeloo, The Netherlands\\
$^2$Kapteyn Institute, Univ. of Groningen, P.O. Box 800, 9700 AV Groningen, The Netherlands}    

\begin{abstract} 
We present first results of new ATCA \HI emission and absorption observations of Centaurus A. The large-scale disk is described via a tilted-ring model. A broader redshifted absorption than previously known as well as a blueshifted absorption component against the nucleus is found.
\end{abstract}

\section{Observations, kinematic modelling and comparison with other models}
We have obtained new ATCA \HI emission and absorption line data of Centaurus A that have better sensitivity and at the same time higher resolution than previous emission-line studies (e.g. van Gorkom et al. 1990) in order to derive a model of the large-scale disk. This will allow to study the fuelling process of the AGN. The \HI disk follows the prominent dust lane and shows a strong warping structure (Fig.~1, left). Absorption is detected against the nucleus, the northern jet and against the southern radio lobe (Fig.~1). The absorption against the nucleus has a larger velocity width than previously known (see Sect.~2). We apply a tilted-ring model directly to the 3-D data cube to derive a model that describes the morphology and kinematics. A model, with the same centre for all rings, describes the complicated warping structure only to first order. For several locations in the data cube the model over- or underpredicts the \Hi . Clearly, this model is not the full answer. We also tried to model the disk using elliptical orbits - but without any success.\\
A comparison with the Spitzer data of Quillen et al. (2006) for the molecular gas shows that our model does not reproduce the Spitzer morphology. On the other hand, Quillen's model does reproduce the \HI morphology (to a satisfying degree) but not the kinematics. Hence, since neither our model nor the Quillen model can reproduce the kinematics satisfyingly (under the assumption of closed orbits), radial motions of the gas must play an important role, showing that the gas disk is not yet relaxed.

\section{The central 120~pc}
Against the nucleus we detect broad abosorption over the range of 400~\kms ~to 800~\kms , which is blue- and redshifted with respect to the systemic velocity (542~\kms ). This absorption is much broader than what was previously known (van der Hulst et al. 1983; Sarma et al. 2002). This is because we now also detect blueshifted absorption, while the redshifted absorption extends to much higher velocities (see Fig. 1, right). All models derived for the large-scale disk so far indicate that a large fraction of the unresolved ($=120$~pc) absorption against the nucleus comes from the nuclear region and is not caused by the large-scale disk being in front of the nucleus in projection. A circum-nuclear \HI disk - perhaps the counterpart of a~$< 2$\arcsec ~disk of molecular and ionised gas (e.g. Marconi et al. 2001) - could explain part of this absorption. However, to explain the full width, additional radial motions (in particular infall) are required. An other possibility is that of a central thick disk, as considered by Eckart et al. (1999). This possibility is still under investigation but so far it does not appear to provide the broad velocity range that we observe.
\begin{figure}
\includegraphics[width=0.52\textwidth]{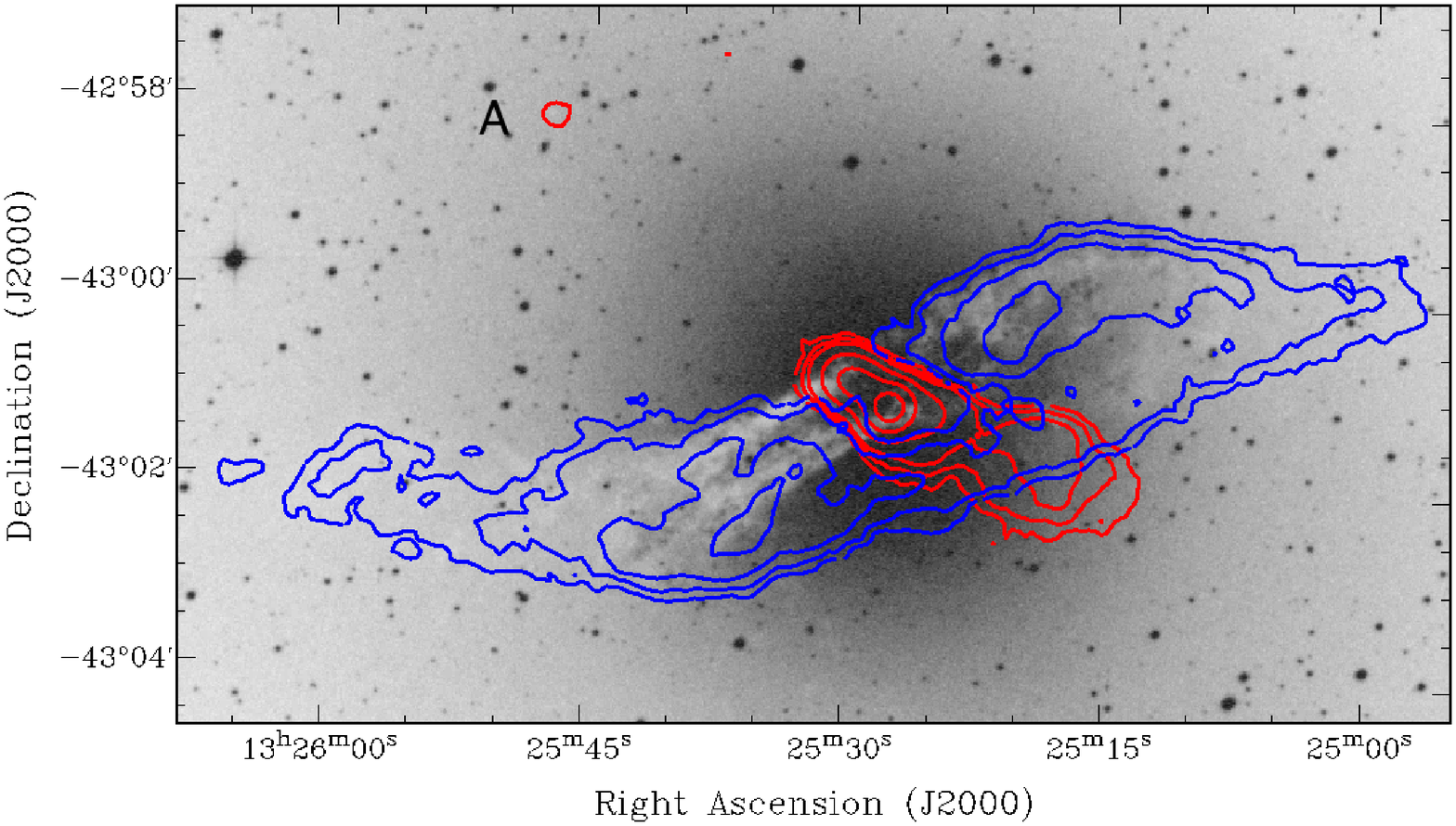}
\includegraphics[width=0.43\textwidth]{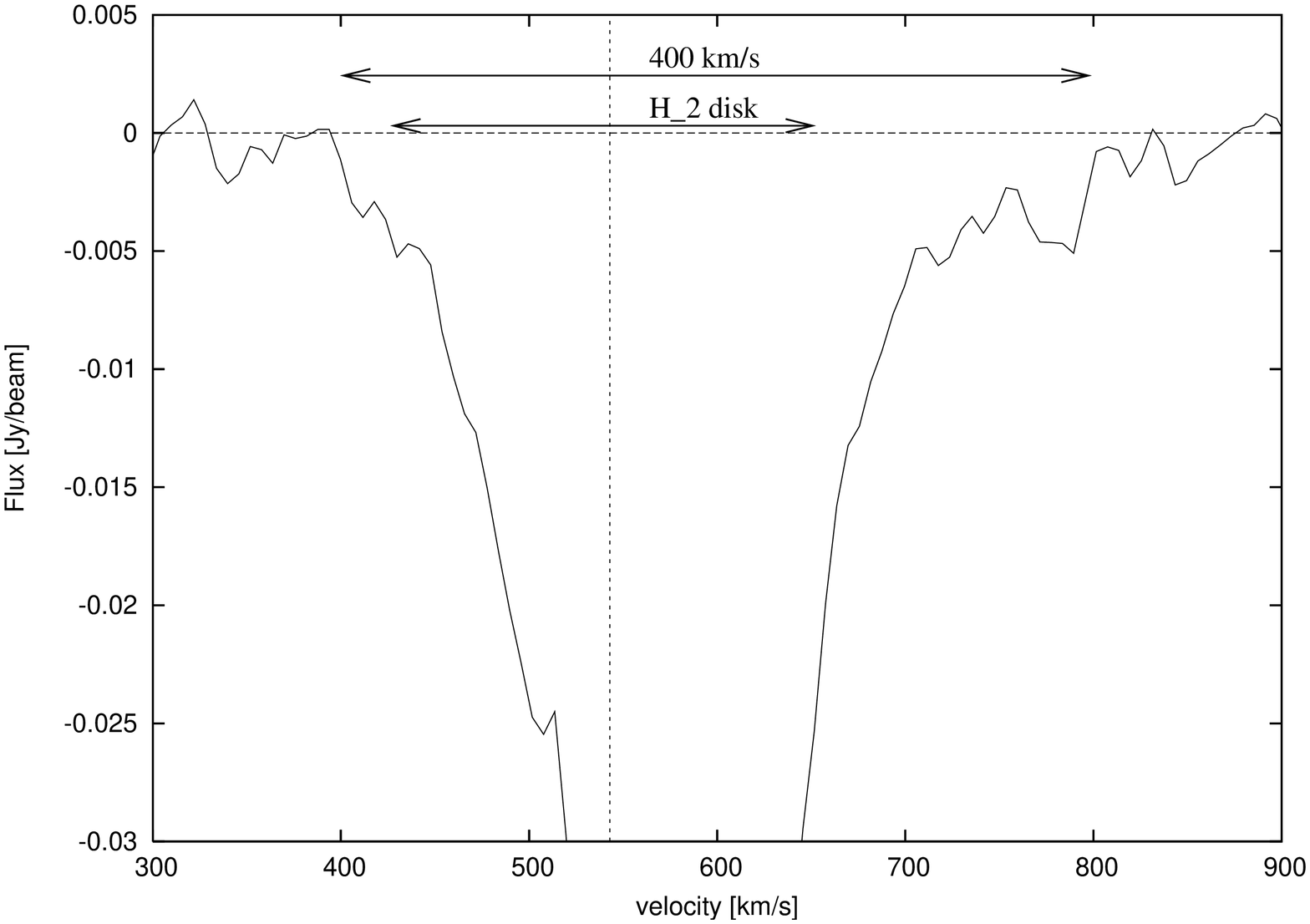}
\caption{Left: DSS2 (red) image overplotted with \HI total intensity contours for emission and absorption. Absorption is detected against the nucleus, the northern jet and southern radio lobe. Contour levels are: -60, -20, -5, -1, -0.5, -0.2 (red) and 0.2, 0.5 1 and 2~Jy/beam.km/s (blue). The red contour labeled with a ``A'' is an absorption cloud. Right: Spectrum of the central 120~pc showing an asymmetry between the red- and blueshifted part of the absorption. The vertical line indicates the systemic velocity. The arrow labeled H$_2$ disk shows the velocity width of the H$_2$ disk found by Marconi et al. (2001).}
\end{figure}

\acknowledgements 
This research was supported by the EU Framework 6 Marie Curie Early Stage
Training programme under contract number MEST-CT-2005-19669 "ESTRELA". The Australia Telescope is funded by the Commonwealth of Australia for operation as a National Facility managed by CSIRO.


\end{document}